\documentclass[12pt]{article}
\usepackage{amsfonts,amsmath,amssymb}
\usepackage{graphicx}

\newfont{\twelvemsb}{msbm10 scaled\magstep1}
\newfont{\eightmsb}{msbm8}
\newfam\msbfam
\textfont\msbfam=\twelvemsb \scriptfont\msbfam=\eightmsb
\catcode`\@=11
\def\Bbb{\ifmmode\let\next\Bbb@\else
\def\next{\errmessage{Use \string\Bbb\space only in math mode}}\fi\next}
\def\Bbb@#1{{\fam\msbfam{{#1}}}}

\newcommand{\be}{\begin{equation}}
\newcommand{\ee}{\end{equation}}
\newcommand{\ba}{\begin{eqnarray}}
\newcommand{\ea}{\end{eqnarray}}

\newcommand{\m}{\mathcal}

\newcommand{\nn}{\nonumber}
\newcommand{\q}{\theta}

\begin{document}

\sloppy
\renewcommand{\thefootnote}{\fnsymbol{footnote}}
\newpage
\setcounter{page}{1} \vspace{0.7cm}
\vspace*{1cm}
\begin{center}
{\bf $\mathcal{N} = 4$ polygonal Wilson loops: fermions.} \\
\vspace{.6cm} {Alfredo Bonini$^{a}$, Davide Fioravanti $^a$, Simone Piscaglia $^{b}$, Marco Rossi $^c$
}
\footnote{E-mail: bonini@bo.infn.it, fioravanti@bo.infn.it, piscaglia@th.phys.titech.ac.jp, rossi@cs.infn.it} \\
\vspace{.3cm} $^a$ {\em Sezione INFN di Bologna, Dipartimento di Fisica e Astronomia,
Universit\`a di Bologna} \\
{\em Via Irnerio 46, 40126 Bologna, Italy}\\

\vspace{.3cm} $^b$ {\em Department of Physics,
Tokyo Institute of Technology,
Tokyo 152-8551, Japan}

\vspace{.3cm} $^c${\em Dipartimento di Fisica dell'Universit\`a
della Calabria and INFN, Gruppo collegato di Cosenza} \\
{\em Arcavacata di Rende, Cosenza, Italy} \\
\end{center}
\renewcommand{\thefootnote}{\arabic{footnote}}
\setcounter{footnote}{0}
\begin{abstract}
{\noindent The contributions of scalars and fermions to the null polygonal bosonic Wilson loops/gluon MHV scattering amplitudes in $\mathcal{N} = 4$ SYM are considered. We first examine the re-summation of scalars at strong coupling. Then, we disentangle the form of the fermion contribution and show its strong coupling expansion. In particular, we derive the leading order with the appearance of a fermion-anti-fermion bound state first and then effective multiple bound states thereof. This reproduces the string minimal area result and also applies to the Nekrasov instanton partition function $\mathcal{Z}$ of the $\mathcal{N}=2$ theories. Especially, in the latter case the method appears to be suitable for a systematic expansion.}
\end{abstract}

\section{Introduction and Summary}
\label{intro}
\setcounter{equation}{0}

${\cal N}=4$\,  Super Yang-Mills (SYM) in the planar limit, with 't Hooft coupling $\lambda = 16 \pi ^2 g^2$, appears at one side (of one example) of the AdS/CFT correspondence \cite{MGKPW1
} and, interestingly, shows remarkable connections with $1+1$ dimensional integrable models \cite{INT
}.
Even if integrability was discovered in the study of anomalous dimensions of local operators, recently techniques borrowed from integrable systems have been used for exact computations of other quantities in the same theory, {\it e.g.} the expectation values of null polygonal (bosonic) Wilson loops (Wls). These Wls are dual to (MHV) gluon scattering amplitudes \cite{AM-amp,Dr1,Dr2,Br,Bern}, which makes them even more interesting, and can be efficiently studied by the all order expansion of the collinear limit of two consecutive edges: their value takes on the form of a (sort of non-local) Operator Product Expansion (OPE) \cite{YSA,BSV1}. In fact, this is the same as the insertion of the identity (operator) as an infinite series of basis states in the space of the integrable quantum GKP string, namely the Form Factor series which sums over the flux-tube excitations: gluons and their bound states, fermions, anti-fermions and scalars.

The validity of the integrable OPE series has been successfully checked, by explicit computations, both in the weak and in the strong coupling regime \cite{Di1,BSV2,Di2,BSV5,Bel1407,Di3,Bel1410,BCCSV1,Bel1501,FPR,BCCSV2,Bel1509,Di4,BFPR1,Di5}. In this letter we shall focus on the latter, whose leading contributions are of the same order and come from two sectors. The first -- due to the non-perturbative string dynamics on $S^5$ --, is computable by considering the scalar excitations \cite{BSV4,BFPR2,BFPR3}; the second one -- caused by the classical string minimal action in $AdS_5$ \cite{AM-amp, TBuA}--, comes from gluons, their bound states and  fermions. As for the scalar series contribution, $W_s$, it is resolutive considering the series for $\ln W_s$: in this manner, each term is proven to be proportional to $\sqrt{\lambda}$. Then, because of the fermion-anti-fermion short range potential (\ref{ferm-antiferm-pot}), they contribute at leading order not as single particles but through a bound state $f\bar{f}$ \cite{BSV3,FPR,BFPR1} which arises only at infinite coupling. Now, the (effective) sum runs on these (free) particles, named 'mesons' ($SU(4)$ singlets). Moreover, it has the same mathematical structure of the Nekrasov instanton partition function $\mathcal{Z}$ of the $\mathcal{N}=2$ theories with $\epsilon_2 \sim 1/g$ \cite{NS}. In fact, there is a short range potential (\ref{pol-pot}) between two mesons which our method uses to produce a systematic expansion at small $\epsilon_2 \sim 1/g$. The leading of the latter is given by a simplified sum on mesons and their multiple bound states which gives rise to the dilogarithm of the Yang-Yang potential, proportional to $\sqrt{\lambda} \propto g \sim1/\epsilon_2$, for the Thermodynamic Bethe Ansatz (TBA). Actually, we have conjectured this kind of TBA contribution in \cite{FPR, BFPR1} on the ground of the scattering theory. In this way we can make a parallel with gluon (stable) bound states and reproduce precisely (the middle node of) the TBA governing the string classical minimal action/area ($=$ free energy) \cite{TBuA,YSA}. In Section \ref {Wilson loops} we briefly describe the contribution of scalars. In Section \ref{fermi}, that of fermions: first, we work out the contribution of $n$ couples $f\bar{f}$ as that of $n$ mesons; then, the sum on (free) mesons (analogues of the instantons in $\mathcal{N}=2$ partition functions) is expanded at small $\epsilon_2 \sim 1/g$. At leading order it becomes the sum on multiple meson bound states which originates the TBA.

\section{Non-Perturbative Scalars in the Wilson Loop}
\label{Wilson loops}

The pentagon OPE approach \cite{BSV1} allows us to represent the Wl as a superposition of pentagonal transitions (squared form factors) and propagations. If we go to the non-perturbative strong coupling regime, scalars decouple themselves to give rise to a relativistic $O(6)$ non-linear $\sigma$-model \cite{AM}. Therefore, we can single out their contribution $W_s$ to the hexagonal Wl  OPE
\be
\label{Wilson}
W_s=\sum_{n=0}^{\infty}W_s^{(2n)} \, , \quad W_s^{(2n)}=\frac{1}{(2n)!}\int\prod_{i=1}^{2n}\frac{d\theta_i}{2\pi}\,G^{(2n)}(\theta_1,\cdots,\theta_{2n})\, e^{- z\sum\limits_{i=1}^{2n}\cosh\theta_i} \, ,
\ee
where only even numbers $2n$ of scalars (with rapidities $\q_i$) are considered, for the Wl/MHV needs to be chargeless under $SU(4)$; the parameter $z=m_{gap}\sqrt{\tau^2+\sigma^2}$ encloses the dependence on two conformal ratios $\sigma$, $\tau $ and is proportional to the dynamically generated mass $m_{gap}(\lambda)$.
Each function $G^{(2n)}$ factorizes $G^{(2n)}=\Pi_{dyn}^{(2n)}\Pi_{mat}^{(2n)}$ into a dynamical factor $\Pi_{dyn}^{(2n)}$, expressed as a product over two-particle functions, and a coupling-independent matrix part\footnote{This factor exhibits an interesting resemblance with the $\mathcal{N}=2$ instanton partition function: in fact, a Young tableaux approach was developed in \cite{BFPR3} to compute $\Pi_{mat}^{(2n)}$.} $\Pi_{mat}^{(2n)}$, encoding the internal $SO(6)$ structure of scalars \cite{BSV4}.
A dramatic improvement occurs when, rather than computing the scalar contribution (\ref{Wilson}), we consider its logarithm
\be\label{logW}
{\cal F}_s=\ln W_s = \sum_{n=1}^{\infty}{\cal F}_s^{(2n)}=\sum_{n=1}^{\infty}\frac{1}{(2n)!}\int\prod_{i=1}^{2n}\frac{d\theta_i}{2\pi}g^{(2n)}(\theta_1,\cdots,\theta_{2n})e^{-z\sum_{i=1}^{2n}\cosh\theta_i}
\ee
by passing from the functions $G^{(2n)}$ to their 'connected' counterparts $g^{(2n)}$, under a customary procedure. The crucial point concerns the {\it asymptotic factorization} of the $G$s: that is to say, when one shifts $2k$ rapidities by a large amount $\Lambda\rightarrow\infty$, while holding fixed the remaining $2n-2k$, $G^{(2n)}$ splits as
\ba\label{factorization}
&& G^{(2n)}(\theta_1+\Lambda,\cdots,\theta_{2k}+\Lambda,\theta_{2k+1},\cdots,\theta_{2n})\ \overset{\Lambda\to\infty}{\longrightarrow}\\
&& G^{(2k)}(\theta_1,\cdots,\theta_{2k})\,G^{(2n-2k)}(\theta_{2k+1},\cdots,\theta_{2n}) + O(\Lambda^{-2})\nn \,.
\ea
This remarkable property crucially affects the connected functions, as
\be
\lim_{\Lambda\to\infty}g^{(2n)}(\theta_1+\Lambda,\cdots,\theta_{m}+\Lambda,\theta_{m+1},\cdots,\theta_{2n})\simeq \frac{1}{\Lambda ^2} \rightarrow 0 \, , \qquad\mbox{for}\ \ m<2n \,, \label {g-asy}
\ee
ensuring eventually their integrability. Clearly, the property (\ref {g-asy}) defines the conformal limit at small $z$ for the logarithm of the Wilson loop,
since, jointly to the relativistic behaviour of the $G^{(2n)}$ (hence the $g^{(2n)}$), it allows us to integrate out one rapidity for each ${\cal F}^{(2n)}_s$ in (\ref{logW}), giving
\be
{\cal F}_s^{(2n)}= \frac{2}{(2\pi)^n (2n)!} \int \prod _{i=1}^{2n-1}d \q_i g^{(2n)}(\q_1, \ldots , \q_{2n-1})  K_0 (z \xi) \,,
\ee
for some known function of the rapidities $\xi(\q_1,\ldots,\q_{2n-1})$ \cite{BFPR2}. Now, we can expand (inside) for small argument the Bessel function
$K_0(z\xi)=-\ln z -\ln\xi + O(1)$ (whilst we could not before with the $G^{(2n)}$). Straightforwardly we can work this out for the leading term and obtain
\be
\ln W_s \simeq A \ln(1/z) \simeq - A \ln m_{gap} \simeq A \frac{\sqrt{\lambda}}{4} \,, \label {fin-scal}
\ee
where the coefficient $A$ is given by a series $A=\sum_{n=1}^{\infty}A^{(2n)}$ over the multi-particle contributions, numerically very convenient as it is rapidly converging \cite{BSV4,Bel1512}. For the sub-leading terms we need a further step as the weak power decay (\ref{g-asy}) compels us to restrict the integral at the region $|z\xi|<1$ and carefully estimate how the rest behaves at small $z$: this is ultimately a consequence of the asymptotic freedom of the $O(6)\ \sigma$-model and gives rise to the peculiar logarithmic behaviour of the two point 2D CFT correlation function \cite{BFPR2, BFPR3}. This procedure can be generalized to higher number of edges and still gives \cite{Fer} a leading term of the form (\ref{fin-scal}), competing with the minimal area term as conjectured in \cite{BSV4}.

\section{Fermion Contribution to the Wilson Loop}
\label{fermi}

We now focus on the contribution to the hexagonal Wilson Loop due to the fermionic sector only: the singlet condition requires $N_f=N_{\bar{f}}$ mod $4$, but in the strong coupling limit only states with $N_f=N_{\bar{f}}$ contribute at the leading order. Anew, the pentagonal OPE writes as a form-factor series
\be\label{WL}
W_f=\sum_{n=0}^{\infty} W_f^{(n)}
\ee
in terms of the contribution of $n$ fermion-anti-fermion couples:
\ba\label{WLferm}
&& W_f^{(n)}=\frac{1}{n! n!}\int _{\m{C}}\prod_{k=1}^n \left[\frac{du_k}{2\pi}\frac{dv_k}{2\pi}
\,\mu_f(u_k)\mu_f(v_k)\,e^{-\tau E_f(u_k)+i\sigma p_f(u_k)}\cdot \right.\\
&& \left. \cdot e^{-\tau E_f(v_k)+i\sigma p_f(v_k)}\right]
\Pi_{dyn}^{(n)}(\{u_i\},\{v_j\})\,\Pi_{mat}^{(n)}(\{u_i\},\{v_j\})
\nn \ .
\ea
The open integration contour $C$, restricted to the small fermion sheet, is described in detail in \cite{BSV3, BFPR1}.
The dynamical quantities are parametrised through the set of fermion $\{u_k\}$ and anti-fermion rapidities $\{v_k\}$: energy and momentum of a particle correspond respectively to $E_f(u)$ and $p_f(u)$ and couple in the propagation phase to the cross ratios $\tau$ and $\sigma$, determining the conformal geometry of the polygon. Analogously to scalars, the multiparticle pentagonal transitions factorize into the product of a dynamical and a (coupling independent) matrix part \cite{BSV4}. The dynamical part in turn is factorized in terms of two particles amplitudes

\small
\be\label{dyn}
\Pi_{dyn}^{(n)}(\{u_i\},\{v_j\}
) = \displaystyle\prod_{i<j}^n\frac{1}{P(u_i|u_j)P(u_j|u_i)}
\frac{1}{P(v_i|v_j)P(v_j|v_i)}\displaystyle\prod_{i,j=1}^n \frac{1}{\bar P(u_i|v_j)\bar P(v_j|u_i)}
\ee
\normalsize
where $P$ stands for the transition between particles of the same type ($i.e.$ fermion-fermion or anti-fermion-anti-fermion) and $\bar{P}$ for the transition between a fermion and an anti-fermion. The function $P(u|v)$ is endowed with a single pole for coinciding rapidities $v=u$, whose residue determines the measure $\mu_f(u)$ \cite{BSV1}: $\mbox{Res}\,_{v=u}\,P(u|v)= i/\mu_f(u)$. The factor $\Pi^{(n)}_{mat}$, encoding the $SU(4)$ matrix structure, has an integral representation \cite{BSV4} in terms of the auxiliary variables $a,\,b,\,c$, corresponding to the nodes of the $SU(4)$ Dynkin diagram. In a system composed of $n$ couples $f\bar{f}$ with rapidities $u_i$, $v_j$, in a $SU(4)$ singlet, the matrix factor reads
\ba\label{Pi_mat^ff}
\Pi_{mat}^{(n)}(\{u_i\},\{v_j\}) &=& \frac{1}{(n!)^3}\int
\prod_{k=1}^n\left(\frac{da_k db_k dc_k}{(2\pi)^3}\right)\cdot \\
&\cdot & \frac{\displaystyle\prod_{i<j}^n g(a_i-a_j) g(b_i-b_j) g(c_i-c_j)}
{\displaystyle\prod_{i,j}^n f(a_i-b_j) f(c_i-b_j) \prod_{i,j}^n f(u_i-a_j) f(v_i-c_j)} \nn \, ,
\ea
where the integrations are performed on the whole real axis and $f(u)=u^2+\frac{1}{4}, \quad g(u)=u^2(u^2+1)$. Similarly to the scalars above \cite{BFPR3}, the multiple integrals (\ref{Pi_mat^ff}) can be evaluated by a Young tableaux method \cite{Fer} and assume, eventually, the polar structure
\be\label{Mat-fer}
\Pi_{mat}^{(n)}(\{u_i\},\{v_j\})=\frac{P^{(n)}(u_1,\dots,u_n,v_1,\dots,v_n)}
{\displaystyle\prod_{i<j}^n[(u_i-u_j)^2+1]\prod_{i<j}^n[(v_i-v_j)^2+1]\prod_{i,j=1}^n[(u_i-v_j)^2+4]} \,.
\ee
$P^{(n)}(u_1,\dots,u_n,v_1,\dots,v_n)$ is a degree $2n(n-1)$ polynomial in the $u_i,\,v_j$\,.

\subsection{Emergence of a bound state}
\label{ffcont}

As we will present in this sub-section, the polar structure of the $SU(4)$ matrix factor (\ref{Mat-fer}) and the properties of the polynomials $P^{(n)}$ play a crucial role to unravel how, in the perturbative strong coupling regime ({\it i.e.} $\lambda\rightarrow\infty$ with the ratios $\bar u_i=u_i/2g$, $\bar v_i=v_i/2g$ finite), the sum on the fermionic sector can be performed as if there is an effective particle, named 'meson', coalescence of a fermion and an anti-fermion. In turn
coalescences of many mesons will be summed up (in the next sub-section) to obtain effectively the right strong coupling limit of the series, in place of the sum over fermions. In this way, we complete the work of \cite{BFPR1}, where only two couples $f\bar{f}$ were analyzed ($n=2$) (cf. also $n=1$ \cite{BSV3}). Actually, already \cite{FPR} conjectured the possibility of substituting the original sum over fermions and anti-fermions with the sum over mesons and their multiple bound states, supposed on the basis of the analytic structure (particle content) of the S-matrix. In details, on the ground of the Bethe Ansatz equations, the meson does not show up in the spectrum at finite coupling, as it lies outside the physical sheet \cite{BSV3,FPR}; on the contrary, it comes into existence at infinitely large values of the coupling and starts contributing to the OPE differently from (unbounded) fermions and anti-fermions, whose contribution is subdominant. The multi-meson bound states share the same destiny \cite{FPR, BFPR1}. To ease our task, we re-cast (\ref{WLferm}) in the form (we could have privileged the $v_j$)
\be\label{Meson}
W_f^{(n)}=\frac{1}{n!}\int_{C}\displaystyle\prod_{i=1}^n\frac{du_i}{2\pi}I_n(u_1,\cdots ,u_n)\displaystyle\prod_{i<j}^n p(u_{ij}) \ ,
\ee
by highlighting a factor accounting for poles and zeroes in the $u_i$ rapidities,
\be
p(u_{ij})=\frac{u_{ij}^2}{u_{ij}^2+1} \, , \quad u_{ij}=u_i-u_j \, ,
\label{pol-pot}
\ee
the (meson-meson) short range potential, and enclosing the integrals on the anti-fermionic rapidities $v_j$ inside the functions

\small
\be\label{I_n}
I_n(u_1, \cdots , u_n)\equiv \frac{1}{n!}\int_{C}\displaystyle\prod_{i=1}^n\frac{dv_i}{2\pi} R_n(\{u_i\},\{v_j\})P^{(n)}(\{u_i\},\{v_j\})\displaystyle\prod_{i,j=1}^n h(u_i-v_j)
\displaystyle\prod_{i<j}^n p(v_{ij}) \ ,
\ee
\normalsize
where we defined the fermion-anti-fermion short range potential \cite{BSV3}
\be
h(u_i-v_j)= \frac{1}{(u_i-v_j)^2+4}
\label{ferm-antiferm-pot} \, .
\ee
$R_n$ is a regular function, with no poles nor zeroes in the rapidities $u_i,\,v_j\,$ and related to the dynamical factor (\ref{dyn}) by
\be\label{R_n}
R_n(\{u_i\},\{v_j\})\displaystyle\prod_{i<j}^n u_{ij}^2v_{ij}^2 \equiv
\Pi_{dyn}^{(n)}(\{u_i\},\{v_j\})\displaystyle\prod_{i=1}^n\hat{\mu}_f(u_i)\hat{\mu}_f(v_i) \ ,
\ee
where the measure and the propagation phase are combined into $\hat{\mu}_f(u) = \mu_f(u)e^{-\tau E_f(u) + i\sigma p_f(u)}$. The strong coupling limit of (\ref{I_n}) can be evaluated by integrating the rapidities $v_i$ by closing the contour $C$ for taking the residues and obtaining the result $I_n^{closed}$. Because of the properties of $P^{(n)}$ \cite{Fer}, only the poles in the fermion-anti-fermion short range potential (\ref{ferm-antiferm-pot}) $v_i=u_j-2i$ survive and provide a contribution to
\be
\label{I_n-clo}
I_n^{closed}(u_1,\cdots ,u_n)=(-1)^n R_n(u_1,\cdots ,u_n, u_1-2i,\cdots ,u_n-2i) \,,
\ee
which means that fermion and anti-fermion pair up to form a complex two-strings with spacing $2i$. A comparison with (\ref{dyn}), (\ref{R_n}) suggests to interpret this two-string (appearing in the OPE) as a bound state particle, the meson, whose energy and momentum are given additively
\be
E_M(u)\equiv E_f(u+i)+E_f(u-i), \quad p_M(u)\equiv p_f(u+i)+p_f(u-i) \ ,
\ee
along with the pentagon transition amplitude built up in the form
\small
\be
P^{MM}(u|v) = -(u-v)(u-v+i) P(u+i|v+i)P(u-i|v-i)|\bar{P}(u-i|v+i)\bar{P}(u+i|v-i) \ .\nn
\ee
\normalsize
From this, we can introduce the regular function (no poles, no zeroes)
\be
P^{MM}_{reg}(u|v)= P^{MM}(u|v) \frac{u-v}{u-v+i}\, \ ,
\ee
for later use and, from $\mbox{Res}\,_{v=u}\,P^{MM}(u|v)= i/\mu_M(u)$, the (hatted) measure
\be
\hat{\mu}_M(u) = \mu_M(u)e^{-\tau E_M(u) + i\sigma p_M(u)}
= -\frac{\hat{\mu}_f(u+i)\hat{\mu}_f(u-i)}{\bar{P}(u+i|u-i)\bar{P}(u-i|u+i)} \ ,
\label{measure}
\ee
which both allow us to recast (\ref{I_n-clo}) in a form with only reference to mesons
\be\label{conj}
I_n^{closed}(u_1,\cdots ,u_n) 
 = \frac{\displaystyle\prod_{i=1}^n\hat{\mu}_M(u_i-i)}{\displaystyle\prod_{i<j}^n P^{MM}_{reg}(u_i-i|u_j-i)P^{MM}_{reg}(u_j-i|u_i-i)} \ .
\ee
Upon plugging this strong coupling limit into (\ref{Meson}), we can efficiently reformulate the fermionic contribution (\ref{WL}) in terms of (free) mesons only:

\small
\ba\label{SingMes}
&& W_f\simeq W^{(M)} = \sum_{n=0}^{\infty}\frac{1}{n!}\int_{C}\displaystyle\prod_{i=1}^n\frac{du_i}{2\pi}\hat{\mu}_M(u_i-i)\cdot \\
&& \cdot\displaystyle\prod_{i<j}^n\frac{1}{P^{MM}_{reg}(u_i-i|u_j-i)
P^{MM}_{reg}(u_j-i|u_i-i)}\displaystyle\prod_{i<j}^n p(u_{ij}) \,.\nn
\ea
\normalsize
Evidently, this expression gives the exact strong coupling limit, though the next orders need a careful reconsideration of the above procedure.


\subsection{Mesons bound states, TBA and beyond}

Now, we shall show that in $W^{(M)}$ (\ref{SingMes}), thanks to the short range potential (\ref{pol-pot}), the sum on mesons may be traded, at leading order, for one on 'TBA effective bound states' (no new nodes for them): this issue reveals a general feature beneath the appearance of a TBA integral equation and a possible physical interpretation of ordinary TBA. Actually, we will develop here a method to go also beyond the leading TBA order, as in and beyond \cite {BF}, in principle at all orders. In fact, formula (\ref{SingMes}) for $W^{(M)}$ shares its form with the instanton partition function $\mathcal{Z}$ of $\mathcal{N}=2$ theories, and from this perspective the large coupling $g\sim 1/\epsilon_2$ for $W^{(M)}$ corresponds to the so-called Nekrasov-Shatashvili limit of $\mathcal{Z}$, where the omega background $\epsilon_2$ approaches zero \cite{NS}. Our approach relies on the introduction of a quantum gaussian field $X(u)$
\be
e^{\left\langle X(u_i)X(u_j)\right\rangle} \equiv \frac{1}{P^{MM}_{reg}(u_i-i|u_j-i)
P^{MM}_{reg}(u_j-i|u_i-i)} \,,
\ee
so that, upon a Hubbard-Stratonovich transformation, we can rewrite the Wl \cite{FPR}
\be
W^{(M)} = \left\langle \sum_{n=0}^{\infty}\frac{1}{n!}\int_{C}\displaystyle\prod_{i=1}^n\frac{du_i}{2\pi }\hat{\mu}_M(u_i-i)e^{X(u_i)}\displaystyle\prod_{i<j}^n p(u_{ij}) \right\rangle \,,
\ee
where the expectation value involves a gaussian path integral over the field $X(u)$ (cf. an analogous development for $\mathcal{Z}$ of $\mathcal{N} = 2$ theories \cite {BF}). Above we have neglected the diagonal terms $u_i=u_j$ of the Gaussian identity as they are sub-leading. The short range potential (\ref{pol-pot}) part can be recast into a determinant form by means of the Cauchy identity
\be
\displaystyle\prod_{i<j}^n p(u_{ij})=\displaystyle\prod_{i<j}^n\frac{u_{ij}^2}{u_{ij}^2+1}=\frac{1}{i^n}\det\left(\frac{1}{u_i-u_j-i}\right) \, .
\ee
Thus, we are encouraged to define the matrix
\be
M(u_i,u_j)\equiv \frac{\left[\hat{\mu}_M(u_i-i)e^{X(u_i)}\hat{\mu}_M(u_j-i)e^{X(u_j)}\right]^{1/2}}{u_i-u_j-i} \,,
\ee
so to obtain the following determinant
\be
W^{(M)} = \left\langle \sum_{n=0}^{\infty}\frac{1}{n!}\int_C\displaystyle\prod_{i=1}^n\frac{du_i}{2\pi i}\det_{ij} M(u_i,u_j) \right \rangle \, .
\ee
In conclusion, this entails the average of a Fredholm determinant
\be\label{WFre}
W^{(M)} = \left\langle\det\left(1 + M\right)\right\rangle = \left\langle \exp\left[\sum_{n=1}^{\infty}\frac{(-1)^{n+1}}{n}\textit{Tr}M^n\right]\right\rangle \,,
\ee
as expanded in the peculiar power traces
\be\label{trace}
\textit{Tr}M^n \equiv \int_{C}\displaystyle\prod_{i=1}^n\frac{du_i}{2\pi i}\hat{\mu}_M(u_i-i)e^{X(u_i)}\displaystyle\prod_{i=1}^n\frac{1}{u_i-u_{i+1}-i}, \quad u_{n+1}\equiv u_1 \, .
\ee
This holds in the same manner for the instanton partition function $\mathcal{Z}$ of $\mathcal{N} = 2$ theories. Now, we need to compute the expansion for large $g\sim 1/\epsilon_2$ of the traces (\ref{trace}). At leading order, we can again close the contour $C$ for $n-1$ rapidities and compute the residues for $u_i-u_{i+1}=i$, obtaining
\be
\textit{Tr}M^n \simeq \frac{(-1)^{n-1}}{n}\int_{C}\frac{du}{2\pi}\hat{\mu}^n_M(u-i)e^{nX(u)}\simeq \frac{(-1)^{n-1}}{n}\int_{C}\frac{du}{2\pi}\hat{\mu}^n_M(u)e^{nX(u)}
\label{traces}
\ee
where the imaginary shifts $\sim 1/g\sim \epsilon_2$ in $\bar u=u/(2 g)$ have been neglected: this is indeed the contribution of a $n$-meson bound state (like for gluons \cite{FPR}). Notice that in $\mathcal{N} = 2$ theories all the integration contours are closed {\it ab initio} \cite{NS}, so that the traces (\ref{trace}) can be, in principle, computed at all orders more easily\cite{Fer}. Instead, for Wls the corrections at next orders have many origins and the computation of the one-loop contribution is much more involved than in \cite{BF}, but here we give a path\cite{Fer}. Within the bound state approximation (\ref{traces}), we can re-sum the Wilson loop (\ref{WFre}) ($\mathcal{N} = 2$ too \cite{NS}) into a simple path integral
\be\label{WLi2}
W^{(M)}\simeq\left\langle\exp\left[-\int_{C}\frac{du}{2\pi}\mu_M(u) Li_2\left[-e^{-\tau E_M(u)+i\sigma p_M(u)} e^{X(u)}\right]\right]\right\rangle \,,
\ee
upon use of (\ref{measure}) (further simplification $\mu_M(u)\simeq -1$). In details, the last gaussian path integral (\ref{WLi2}) can be re-interpreted as the partition function with an effective action, Yang-Yang potential, with dilogarithm potential and coincides with the conjecture of \cite{FPR} for the middle node of the $A_3$ TBA \cite{TBuA}: the stationary point of the Yang-Yang potential gives the TBA equations. In fact, the other two nodes TBA contributions to the effective action can be obtained by summing up the contribution of the two (components of the) gluons, which genuinely form bound states (and then the dilogarithm potential \cite{FPR}). Of course, the saddle point TBA equations are indeed the leading order since the effective action is proportional to $g$; moreover, they coincide with those arising, in a fully different manner, by minimizing the string area/action. The whole procedure of this section in two steps, -- emergence of meson and effectiveness of its bound states --, extends to all the other polygons thus opening the way to the treatment of \cite{BFPR1}.

\section{Conclusions and Perspectives}
\label{conc}

For scalars and fermions we compute the coupling independent parts of the OPE series as some random partitions on Young tableaux. This allows us to disentangle their respective two contributions (of the same order) at large coupling. At infinite coupling, fermion-anti-fermion pairs have been thought of as mesons which, by virtue of the short range potential (\ref{pol-pot}), form bound states namely generate the $1/n$ factor (in the traces (\ref{traces})) which yields the typical TBA (di)logarithm form. Importantly, the method is amenable to give a systematic expansion also for the partition function $\mathcal{Z}$ of $\mathcal{N} = 2$ gauge theories, with instanton positions $u_i$ (and their bound states at leading order)\cite{Fer}.

\medskip
{\bf Acknowledgements}
Supporting grants: GAST (from INFN), EC Network Gatis and JSPS Postdoctoral Fellows 16F16735 for S.P..


\begin{thebibliography}{99.}


\bibitem{TBuA}
L. Alday, D. Gaiotto, J. Maldacena,
JHEP{\bf 09} (2011) 32, arXiv:0911.4708 [hep-th];


\bibitem{AM-amp}
L. Alday, J. Maldacena,
JHEP{\bf 06} (2007) 064, arXiv:0705.0303 [hep-th];

\bibitem{AM}
L. Alday, J. Maldacena,
JHEP{\bf 11} (2007) 019, arXiv:0708.0672 [hep-th];

\bibitem{YSA}
L. Alday, J. Maldacena, A. Sever, P. Vieira,
J. Phys. A {\bf 43} 485401 (2010), arXiv:1002.2459 [hep-th];

\bibitem{BCCSV1}
B. Basso, J. Caetano, L. Cordova, A. Sever, P. Vieira,
JHEP{\bf 08} (2015) 018 and arXiv:1412.1132 [hep-th];

\bibitem{BCCSV2}
B. Basso, J. Caetano, L. Cordova, A. Sever, P. Vieira,
JHEP{\bf 12} (2015) 01 and arXiv:1508.02987 [hep-th];

\bibitem{BSV1}
B. Basso, A. Sever, P. Vieira,
Phys. Rev. Lett. {\bf 111} (2013) 091602, arXiv:1303.1396 [hep-th];

\bibitem{BSV2}
B. Basso, A. Sever, P. Vieira,
JHEP{\bf 01} (2014) 008 and arXiv:1306.2058 [hep-th];

\bibitem{BSV3}
B. Basso, A. Sever, P. Vieira,
JHEP{\bf 08} (2014) 085, arXiv:1402.3307 [hep-th];

\bibitem{BSV4}
B. Basso, A. Sever, P. Vieira,
Phys. Rev. Lett. {\bf 113} (2014) 26, 261604, arXiv:1405.6350 [hep-th];

\bibitem{BSV5}
B. Basso, A. Sever, P. Vieira,
JHEP{\bf 09} (2014) 149 and arXiv:1407.1736 [hep-th];

\bibitem{INT}
N. Beisert et al., Lett. Math. Phys. {\bf 99} (2012) 3, arXiv:1012.3982 [hep-th];

\bibitem{Bel1407}
A.V. Belitsky, Nucl. Phys. B {\bf 896} (2015) 493, arXiv:1407.2853 [hep-th];

\bibitem{Bel1410}
A.V. Belitsky, Nucl. Phys. B {\bf 894} (2015) 108, arXiv:1410.2534 [hep-th];

\bibitem{Bel1501}
A.V. Belitsky, Nucl. Phys. B {\bf 897} (2015) 346, arXiv:1501.06860 [hep-th];

\bibitem{Bel1509}
A.V. Belitsky, Nucl. Phys. B {\bf 911} (2016) 517, arXiv:1509.06054 [hep-th];

\bibitem{Bel1512}
A.V. Belitsky, Nucl. Phys. B {\bf 911} (2016) 425, arXiv:1512.00555 [hep-th];

\bibitem{Bern}
Z. Bern, L. J. Dixon, D. A. Kosower, R. Roiban, M. Spradlin, C. Vergu, A. Volovich,  Phys. Rev. D {\bf 78} (2008) 045007,  arXiv:0803.1465 [hep-th];

\bibitem{BFPR1}
A. Bonini, D. Fioravanti, S. Piscaglia, M. Rossi,
JHEP{\bf 04} (2016) 029, arXiv:1511.05851 [hep-th];

\bibitem{BFPR2}
A. Bonini, D. Fioravanti, S. Piscaglia, M. Rossi,
Phys.Rev. D {\bf 95} (2017) no.4, 041902, arXiv:1607.02084 [hep-th];

\bibitem{BFPR3}
A. Bonini, D. Fioravanti, S. Piscaglia, M. Rossi,
arXiv:1707.05767 [hep-th];


\bibitem{Fer}
A. Bonini, D. Fioravanti, S. Piscaglia, M. Rossi, longer, detailed paper to appear;

\bibitem{BF}
J.E. Bourgine, D. Fioravanti, Phys. Lett. B {\bf 750} 139, arXiv:1506.01340 [hep-th];

\bibitem{Br}
A. Brandhuber, P. Heslop and G. Travaglini, Nucl. Phys. B {\bf 794} (2008) 231, arXiv 0707.1153 [hep-th];

\bibitem{Di5}
S. Caron-Huot, L. Dixon, A. McLeod, M. von Hippel, Phys. Rev. Lett. {\bf 117} (2016) 241601, arXiv:1609.00669 [hep-th];

\bibitem{Di1}
L. Dixon, J. Drummond, J. Henn, JHEP {\bf 1111} (2011) 023, arXiv:1108.4461 [hep-th];

\bibitem{Di2}	
L. Dixon, J. Drummond, M. von Hippel, J. Pennington, JHEP {\bf 1312} (2013) 049, arXiv:1308.2276 [hep-th];

\bibitem{Di3}
L. Dixon, M. von Hippel, JHEP {\bf 1410} (2014) 065, arXiv:1408.1505 [hep-th];

\bibitem{Di4}
L. Dixon, M. von Hippel, A. McLeod, JHEP {\bf 1601} (2016) 053, arXiv:1509.08127 [hep-th];

\bibitem{Dr1}
J.M. Drummond, J. Henn, G.P. Korchemsky, E. Sokatchev, Nucl. Phys. B {\bf 795} (2008) 52, arXiv: 0709.2368 [hep-th];

\bibitem{Dr2}
J.M. Drummond, J. Henn, G.P. Korchemsky, E. Sokatchev, Nucl. Phys. B {\bf 815} (2009) 142, arXiv: 0803.1466 [hep-th];


\bibitem{FPR}
D. Fioravanti, S. Piscaglia, M. Rossi,
Nucl. Phys. B {\bf 898} (2015) 301, arXiv:1503.08795 [hep-th];


\bibitem{MGKPW1}
J. Maldacena,
Adv. Theor. Math. Phys. {\bf 2} (1998) 231, hep-th/9711200;


\bibitem{NS}
N. Nekrasov, S. Shatashvili, arXiv:0908.4052[hep-th], Proc.16th C.Int.Math.Phy.



\end{thebibliography}
\end{document}